\documentclass[twocolumn,showpacs,aps,prl,superscriptaddress,letterpaper]{revtex4}
\usepackage{graphicx}
\usepackage{dcolumn}
\usepackage{amsmath}
\usepackage{epsfig}
\usepackage{multirow}

\long\def\inst#1{\par\nobreak\kern 4pt\nobreak
    {\itshape #1}\par\vskip 10pt plus 3pt minus 3pt}

\RequirePackage{xspace}
\usepackage{relsize}

\def\babar{\mbox{\slshape B\kern-0.1em{\smaller A}\kern-0.1em
    B\kern-0.1em{\smaller A\kern-0.2em R}}}
\def\Abar    {\kern 0.18em\overline{\kern -0.18em A}{}\xspace}
\def\Kbar    {\kern 0.18em\overline{\kern -0.18em K}{}\xspace}
\def\Dbar    {\kern 0.18em\overline{\kern -0.18em D}{}\xspace}
\def\Bbar    {\kern 0.18em\overline{\kern -0.18em B}{}\xspace}

\def\BB      {\ensuremath{B\Bbar}\xspace} 
\def\Bz      {\ensuremath{B^0}\xspace}
\def\Bzb     {\ensuremath{\Bbar^0}\xspace}
\def\BzBzb   {\ensuremath{\Bz {\kern -0.16em \Bzb}}\xspace}
\def\Bu      {\ensuremath{B^+}\xspace}
\def\Bub     {\ensuremath{B^-}\xspace}

\def\BpBm    {\ensuremath{\Bu {\kern -0.16em \Bub}}\xspace}

\newcommand{\optbar}[1]{\shortstack{{\tiny (\rule[.4ex]{1em}{.1mm})}
  \\ [-.7ex] $#1$}}
\def\BorBbar    {\kern 0.18em\optbar{\kern -0.18em B}{}\xspace}
\def\DorDbar    {\kern 0.18em\optbar{\kern -0.18em D}{}\xspace}
\def\KorKbar    {\kern 0.18em\optbar{\kern -0.18em K}{}\xspace}
\def\CP                {\ensuremath{C\!P}\xspace}
\def\pep2{PEP-II}
\mathchardef\Upsilon="7107
\def\Y#1S{\ensuremath{\Upsilon{(#1S)}}\xspace}

\def\FourS {\Y4S}

\newcommand{\gevcc}{\ensuremath{{\mathrm{\,Ge\kern -0.1em V\!/}c^2}}\xspace}
\newcommand{\gev}{\ensuremath{\mathrm{\,Ge\kern -0.1em V}}\xspace}

\def\mes        {\mbox{$m_{\rm ES}$}\xspace}
\def\DeltaE     {\mbox{$\Delta E$}\xspace}

\newcommand{\mev}{\ensuremath{\mathrm{\,Me\kern -0.1em V}}\xspace}
\newcommand{\mevcc}{\ensuremath{{\mathrm{\,Me\kern -0.1em V\!/}c^2}}\xspace}

\newcommand{\BABARPubYear}     {08}
\newcommand{\BABARPubNumber}  {021}

\newcommand{\SLACPubNumber} {13267}

\begin{document}

\begin{flushleft}
\babar-PUB-\BABARPubYear/\BABARPubNumber\\
SLAC-PUB-\SLACPubNumber\\
\end{flushleft}

\title{\boldmath 
Observation and Polarization Measurements of $B^\pm\to\varphi K_1^\pm$
and $B^\pm\to\varphi K_2^{*\pm}$
}

%
\author{B.~Aubert}
\author{M.~Bona}
\author{Y.~Karyotakis}
\author{J.~P.~Lees}
\author{V.~Poireau}
\author{E.~Prencipe}
\author{X.~Prudent}
\author{V.~Tisserand}
\affiliation{Laboratoire de Physique des Particules, IN2P3/CNRS et Universit\'e de Savoie, F-74941 Annecy-Le-Vieux, France }
\author{J.~Garra~Tico}
\author{E.~Grauges}
\affiliation{Universitat de Barcelona, Facultat de Fisica, Departament ECM, E-08028 Barcelona, Spain }
% ...
\collaboration{The \babar\ Collaboration}
\noaffiliation
\date{October 22, 2008}
\begin{abstract}
With the full $\babar$ data sample of $465~\times 10^6$ $B\Bbar$ pairs, 
we observe the decays $B^\pm\to\varphi K_1(1270)^\pm$ 
and $B^\pm\to\varphi K_2^{*}(1430)^{\pm}$.
We measure the branching fractions $(6.1\pm{1.6}\pm1.1)\times10^{-6}$ and 
$(8.4\pm 1.8\pm1.0)\times10^{-6}$ and the fractions of longitudinal polarization
$0.46^{+0.12+0.06}_{-0.13-0.07}$ and $0.80^{+0.09}_{-0.10}\pm0.03$, respectively.
We also report on the $B^\pm\to\varphi K_0^{*}(1430)^{\pm}$ decay branching 
fraction of $(7.0\pm1.3\pm0.9)\times10^{-6}$ and several parameters sensitive
to $C\!P$ violation and interference in the above three decays.
Upper limits are placed on the $B^{\pm}$ decay rates to final states with 
$\varphi$ and $K_1(1400)^{\pm}$, $K^*(1410)^{\pm}$, $K_2(1770)^{\pm}$, 
or $K_2(1820)^{\pm}$. 
Understanding the observed polarization pattern requires 
amplitude contributions from an uncertain source.
\end{abstract}

\maketitle

Measurements of polarization in rare vector-vector $B$ meson decay, 
such as $B\to\varphi K^*$~\cite{babar:vv,belle:phikst}, 
have revealed an unexpectedly large fraction of transverse polarization
and suggested contributions to the decay amplitude which 
were previously neglected. 
Decays to other excited spin-$J$ kaons $K_J^{(*)}$ can also take place.
The differential width for a $B\to\varphi K_J^{(*)}$ decay 
has three complex amplitudes $A_{J\lambda}$, which describe
the three helicity states $\lambda=0,\pm1$, except when $J=0$.
The expected hierarchy of the $A_{J\lambda}$ amplitudes 
$|A_{J0}|^2\gg |A_{J+}|^2\gg |A_{J-}|^2$ is sensitive to the $(V-A)$ structure 
of the weak interactions with the left-handed fermion 
couplings~\cite{smtheory, bvvreview2006, bib:Amsler2008},
and therefore is sensitive to physics beyond the standard model.
For example, tensor or scalar interactions would violate $|A_{J0}|^2\gg |A_{J+}|^2$
and the right-handed fermion couplings would violate 
$|A_{J+}|^2\gg |A_{J-}|^2$~\cite{smtheory}. Strong interaction effects
could change these predictions as well, but were originally expected 
to be small~\cite{smtheory}.

However, all previous studies have been limited to the two-body $K^*_J\to K\pi$ 
decays, thus considering only the spin-parity $K^*_J$ states with $P=(-1)^{J}$.
In this paper we report the measurement with the three-body final states 
$K_J^{(*)}\to K\pi\pi$ which include $P=(-1)^{J+1}$ mesons such as $K_1$ and $K_2$.
We complement these measurements with the two-body $K^{(*)}_J$ final states
in the $B^\pm$ decays and report polarization in the 
$\varphi K_1(1270)^\pm$ and $\varphi K_2^{*}(1430)^{\pm}$
final states which have not been seen before.
We also search for other final states with $\varphi$ 
and $K_0^*(1430)^{\pm}$, $K_1(1400)^{\pm}$, $K^*(1410)^{\pm}$, 
$K_2(1770)^{\pm}$, or $K_2(1820)^{\pm}$. 


We use data collected with the $\babar$ detector~\cite{babar} at the 
PEP-II $e^+e^-$ collider. A sample of $(465\pm 5)\times 10^6$ 
$\FourS\to\BB$ events was recorded at the 
the $e^+e^-$ center-of-mass energy $\sqrt{s} = 10.58\gev$.
Momenta of charged particles are measured 
in a tracking system consisting of a silicon vertex tracker with five 
double-sided layers and a 40-layer drift chamber, both within the 1.5-T 
magnetic field of a solenoid. 
Identification of charged particles is provided 
by measurements of the energy loss in the tracking devices and by 
a ring-imaging Cherenkov detector. 
Photons are detected by a CsI(Tl) electromagnetic calorimeter.

We search for $B^\pm\to\varphi K_J^{(*)\pm}$ decays using three 
final states of the $K_J^{(*)\pm}$ decay: $K^0_S\pi^\pm$, $K^\pm\pi^0$,
and $K^\pm\pi^+\pi^-$, where $K^0_S\to\pi^+\pi^-$ and $\pi^0\to\gamma\gamma$.
We define the two helicity angles $\theta_i$ as the angle between the direction
of the $K$ or $K^+$ meson from $K^*\to K\pi$ ($\theta_1$) or 
$\varphi\to K^+K^-$ ($\theta_2$) and the direction opposite to the 
$B$ in the $K^*$ or $\varphi$ rest frame. 
The normal to the three-body decay plane for $K_J^{(*)}\to K\pi\pi$
is chosen as the analyzer of the $K_J^{(*)}$ polarization instead 
of the direction of $K$ from $K_J^*$ in the two-body decays. 
We define ${\cal H}_i=\cos\theta_i$.

We identify $B$ meson candidates using two kinematic variables:
$\mes = (s/4-\mathbf{p}_B^2)^{1/2}$
and $\DeltaE = \sqrt{s}/2 - E_B$,
where $(E_B,\mathbf{p}_B)$ is the four-momentum of the $B$ candidate
in the $e^+e^-$ center-of-mass frame.
We require $m_{\rm{ES}}>5.25$ GeV and 
$|\Delta{E}|<0.1$ GeV (or 0.08 GeV for $K_J^{(*)\pm}\to K^\pm\pi^+\pi^-$) GeV.
We also require the invariant masses to satisfy
$1.1<m_{K\pi}<1.6$ GeV,
$1.1<m_{K\pi\pi}<2.1$ GeV, and
$0.99<m_{K^+K^-}<1.05$ GeV.
To reject the dominant $e^+e^-\to$ light quark-antiquark 
background, we use the angle $\theta_T$ between
the thrust axis of the $B$-candidate decay products
and that of the rest of the event
requiring $|\cos\theta_T|<0.8$, 
and a Fisher discriminant ${\cal F}$ 
which combines event-shape parameters~\cite{bigPRD}.

To reduce combinatorial background in the mode 
$K_J^{*\pm}\to K^\pm\pi^0$, we require  ${\cal H}_1<0.6$.
When more than one candidate is reconstructed
($7.6\%$ of events with $K^{0}_{S}\pi^\pm$, $2.9\%$ with $K^\pm\pi^0$, 
and $14.6\%$ with $K^\pm\pi^+\pi^-$),
we select the one whose $\chi^2$ of the charged-track vertex fit 
combined with $\chi^2$ of the invariant mass consistency of 
the $K^0_S$ or $\pi^0$ candidate, is the lowest. 
We define the $b$-quark flavor sign $Q$ to be opposite 
to the charge of the $B$ meson candidate.


We use an unbinned extended maximum-likelihood fit~\cite{babar:vv}
to extract the event yields $n_{j}$ and the probability 
density function (PDF) parameters,  denoted by
{\boldmath$\zeta$} and {\boldmath$\xi$}, to be described below.
The index $j$ represents the event categories, 
which include continuum background and several $B$-decay modes.
In the $B^\pm\to\varphi K_J^{*\pm}\to(K^+K^-)(K\pi)$ topology, 
the following event categories are considered:
$\varphi K_2^{*}(1430)^{\pm}$,
$\varphi (K\pi)^{*\pm}_0$, and
$f_0 (K\pi)^{*\pm}_0$, where the $J^P=0^+$  $(K\pi)^{*\pm}_0$
contribution includes both a nonresonant component 
and the $K_0^{*}(1430)^\pm$ resonance~\cite{Aston:1987ir}.
In the $B^\pm\to\varphi K_J^{(*)\pm}\to(K^+K^-)(K\pi\pi)$ 
topology, we consider $\varphi K_1(1270)^{\pm}$, 
$\varphi K_1(1400)^{\pm}$, 
$\varphi K_2^{*}(1430)^{\pm}$,
$\varphi K^*(1410)^{\pm}$, 
$\varphi K_2(1820)^{\pm}$,
a nonresonant $\varphi K^\pm\pi^+\pi^-$, and
$f_0 K_1(1400)^{\pm}$ contributions. 
In the latter topology, the mode $\varphi K_2(1770)^{\pm}$ is also
considered in place of $\varphi K_2(1820)^{\pm}$.
In all cases, the modes with $f_0$ model
can account for a possible broad non-$\varphi$ 
 $(K^+K^-)$ contributions under the $\varphi$.

The extended likelihood is
${\cal L} = \exp\left(-\sum n_{j}\right)\,\prod{{\cal L}_i}$.
The likelihood ${\cal L}_i$ for candidate $i$ is defined as
${\cal L}_i = \sum_{j,k}n_{j}^k\, 
{\cal P}_{j}^k$({\boldmath ${\rm x}_i$};~{\boldmath$\zeta$},~{\boldmath$\xi$}),
where ${\cal P}_{j}^k$ is the PDF for variables
{\boldmath ${\rm x}_i$}~$=\{{\cal H}_1$, ${\cal H}_2$,  
$m_{K\!\pi(\pi)}$, $m_{K^+K^-}$, $\Delta E$, $m_{\rm{ES}}$, ${\cal F}$, $Q$\}.
The flavor index $k$ corresponds to the value of $Q$, that is
${\cal P}_{j}^k\equiv{\cal P}_{j}\times\delta_{kQ}$. 
The {\boldmath$\zeta$} are the polarization parameters, 
only relevant for the signal PDF.
The {\boldmath$\xi$} parameters describe the background or the 
remaining signal PDFs, which 
are left free to vary in the fit for 
the combinatorial background and are fixed to the values extracted from 
Monte Carlo (MC) simulation~\cite{geant} and calibration 
$B\to\Dbar\pi$ decays in other cases.

The signal PDF for a given candidate $i$ is a joint PDF for 
the helicity angles and resonance mass, and the product of 
the PDFs for each of the remaining variables. 
The helicity part of the signal PDF is the 
ideal angular distribution from Ref.~\cite{vtpaper},
multiplied by an empirical acceptance function
${\cal{G}}({\cal H}_1,{\cal H}_2)$. 
In the $B\to\varphi K_1$ or $\varphi K_2$ parameterization, 
the additional kinematic parameters for the decays 
$K_J^{\pm}\to K^{\pm}\pi^+\pi^-$ (such as $r_1$, $r_2$, and 
$r_{02}$ in Ref. \cite{vtpaper}) are modeled using the sequential 
two-body decay chains~\cite{bib:Amsler2008}.
A relativistic spin-$J$ Breit--Wigner amplitude parameterization
is used for the resonance masses~\cite{bib:Amsler2008,f0mass}, and the
$J^P=0^+$ $(K\pi)^{*\pm}_0$ $m_{K\!\pi}$ amplitude is parameterized 
with the LASS function~\cite{Aston:1987ir}.
The nonresonant $\varphi K^\pm\pi^+\pi^-$ contribution is 
modeled through sequential $K^*(892)\pi\to K\pi\pi$ decay,
while the decay $K\rho\to K\pi\pi$ is considered in the
systematic uncertainty studies.
We use a sum of Gaussian functions 
for the parameterization of $\Delta E$, $m_{\rm{ES}}$, and ${\cal F}$.

The interference between the $J=2$ and $0$ $(K\pi)^{\pm}$ contributions 
is modeled with the term $2{\cal R\rm e}(A_{20} A^*_{00})$, 
with the three-dimensional angular and $m_{K\!\pi}$ 
parameterization. 
We allow an unconstrained flavor-dependent overall shift
$(\delta_0+\Delta\delta_0\times Q)$ between the LASS amplitude 
phase and the tensor resonance amplitude phase.
The polarization parameters {\boldmath$\zeta$} include
the fractions of longitudinal polarization
$f_L={|A_{J0}|^2/\Sigma|A_{J\lambda}|^2}$
in several channels, $\delta_0$, and $\Delta\delta_0$.
Similar interference between the $K_1(1270)^{\pm}$ and $K_1(1400)^{\pm}$
contributions is allowed in the study of systematic uncertainties
but is not included in the nominal fit due to observed dominance 
of only one mode and therefore unconstrained phase of the interference.

Since the $K_2^{*}(1430)^{\pm}$ meson contributes to all three
$K^0\pi^\pm$, $K^\pm\pi^0$, and $K^\pm\pi^+\pi^-$ final states and 
$(K\pi)_0^{*\pm}$ contributes to two $K\pi$ final states in this analysis, 
we consider the total ${\cal L}$ as a product of three likelihoods
constructed for each of the three channels.
The corresponding yields in different channels are related 
by the relative efficiency.
We fit the yields in each charge category $k$ independently
and report them in the form of the total yield $n_j=n_j^++n_j^-$ and 
direct-$C\!P$ asymmetry ${\cal A}_{C\!P}=(n_j^+-n_j^-)/n_j$. 

The combinatorial background PDF is the product of the 
PDFs for independent variables and is found to describe 
well both the dominant quark-antiquark background and the 
background from random combinations of $B$ tracks.
We use polynomials for the PDFs, 
except for $m_{\rm{ES}}$ and ${\cal F}$ distributions
which are parameterized by an empirical phase-space 
function and by Gaussian functions, respectively.
Resonance production occurs in the background and 
is taken into account in the PDF.


\begin{figure}[t]
\centerline{
\setlength{\epsfxsize}{0.5\linewidth}\leavevmode\epsfbox{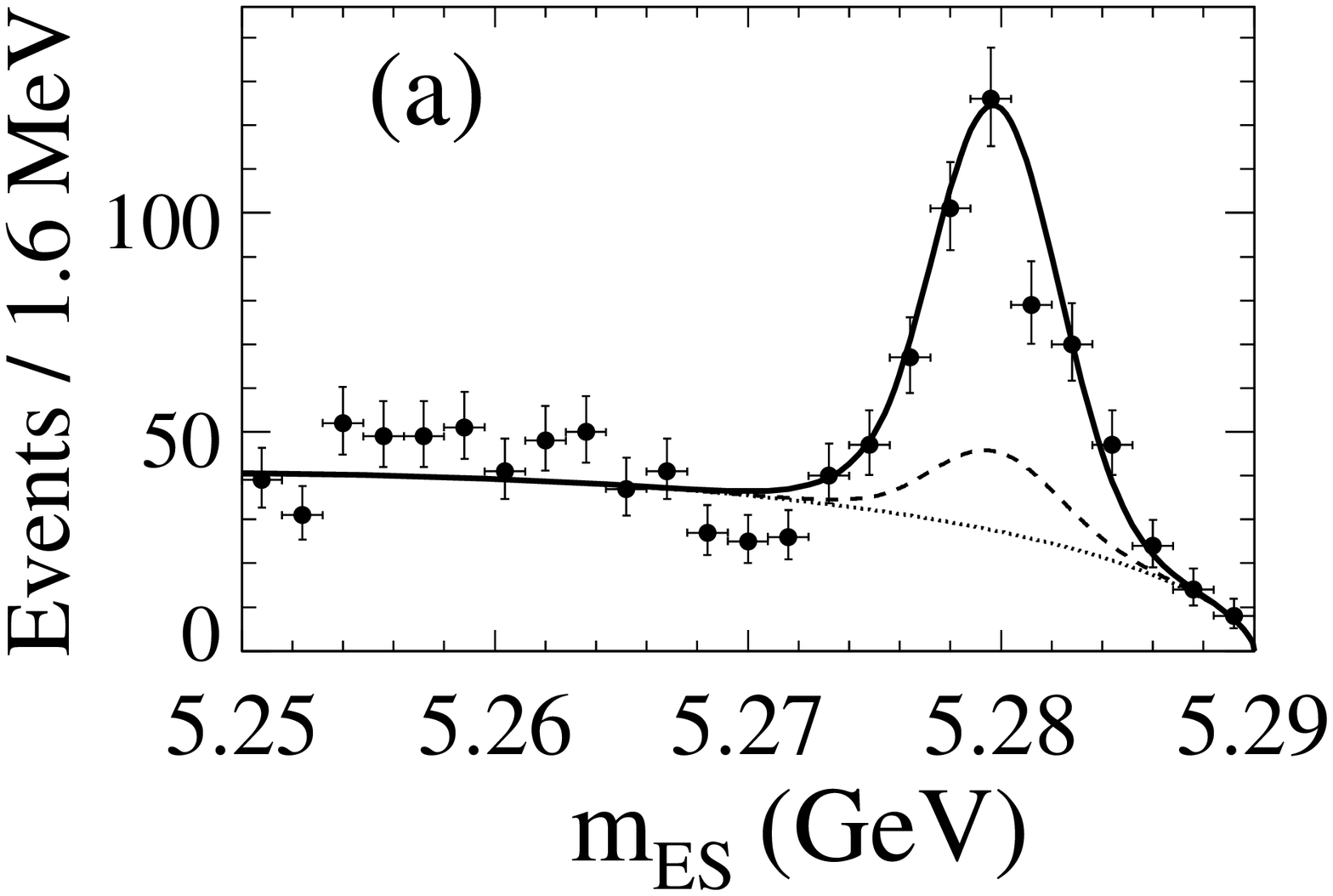}
\setlength{\epsfxsize}{0.5\linewidth}\leavevmode\epsfbox{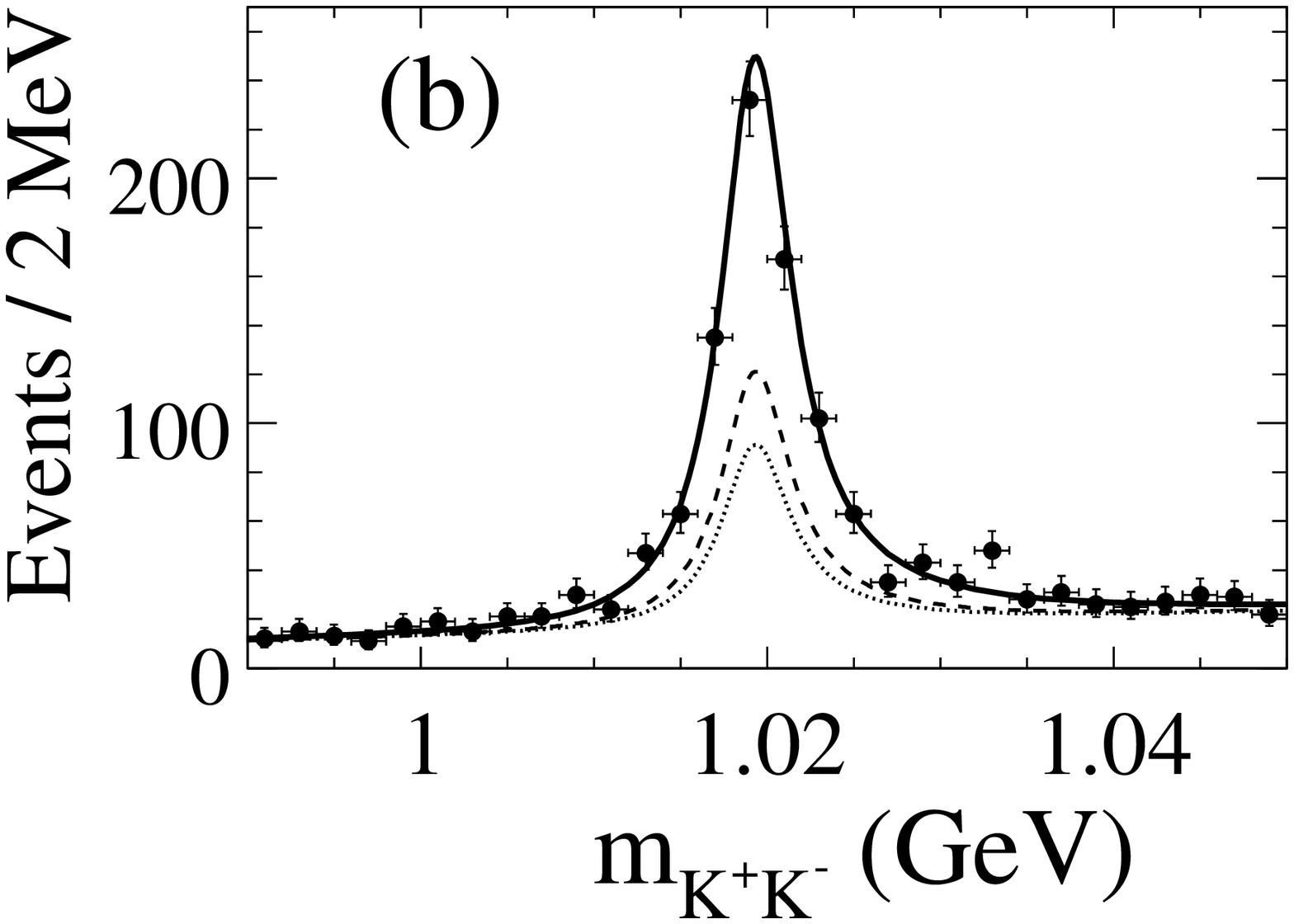}
}
\vspace{-0.3cm}
\caption{\label{fig:projection1} 
Projections onto the variables  $m_{\rm ES}$ (a), and  $m_{K\!\Kbar}$ (b) 
for the signal $B^+\to\varphi(K\pi)$ and $B^+\to\varphi(K\pi\pi)$ candidates.
Data distributions are shown with a requirement on the signal-to-background
probability ratio calculated with the plotted variable excluded.
The solid (dotted) lines show the signal-plus-background
(combinatorial background) PDF projections, while the dashed lines show the 
full PDF projections excluding the signal.
}
\end{figure}


\begin{figure}[t]
\centerline{
\setlength{\epsfxsize}{0.5\linewidth}\leavevmode\epsfbox{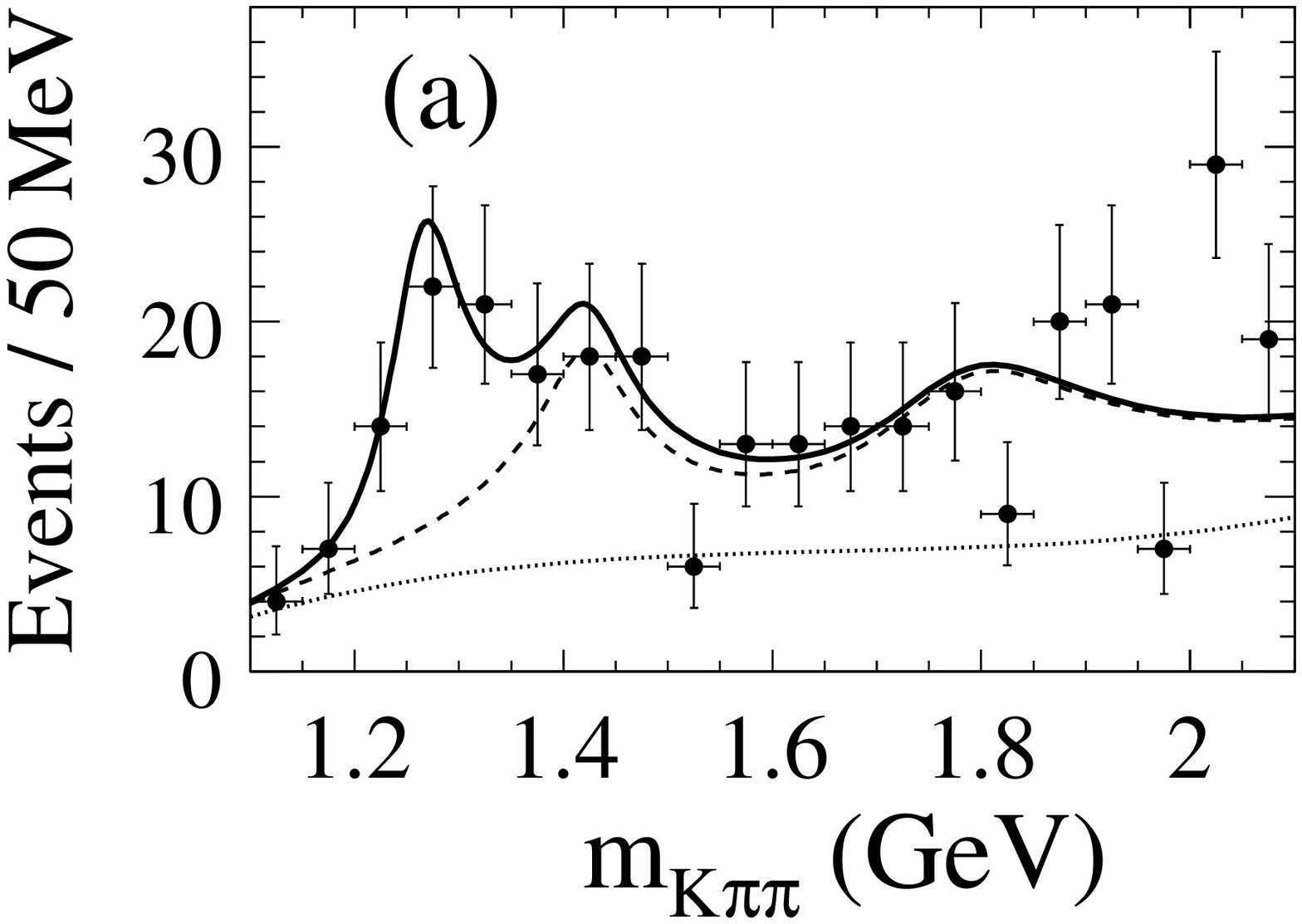}
\setlength{\epsfxsize}{0.5\linewidth}\leavevmode\epsfbox{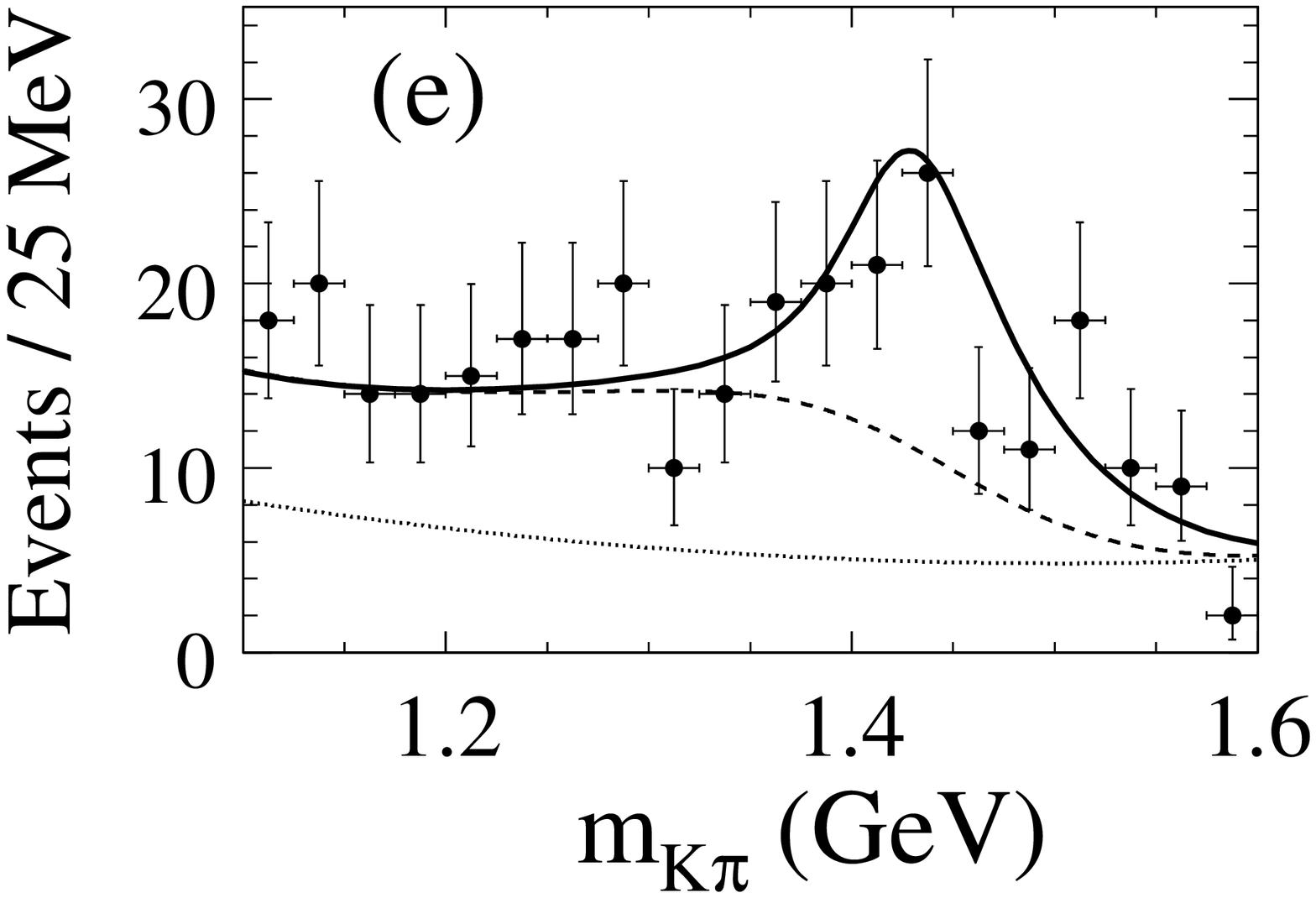}
}
\centerline{
\setlength{\epsfxsize}{0.5\linewidth}\leavevmode\epsfbox{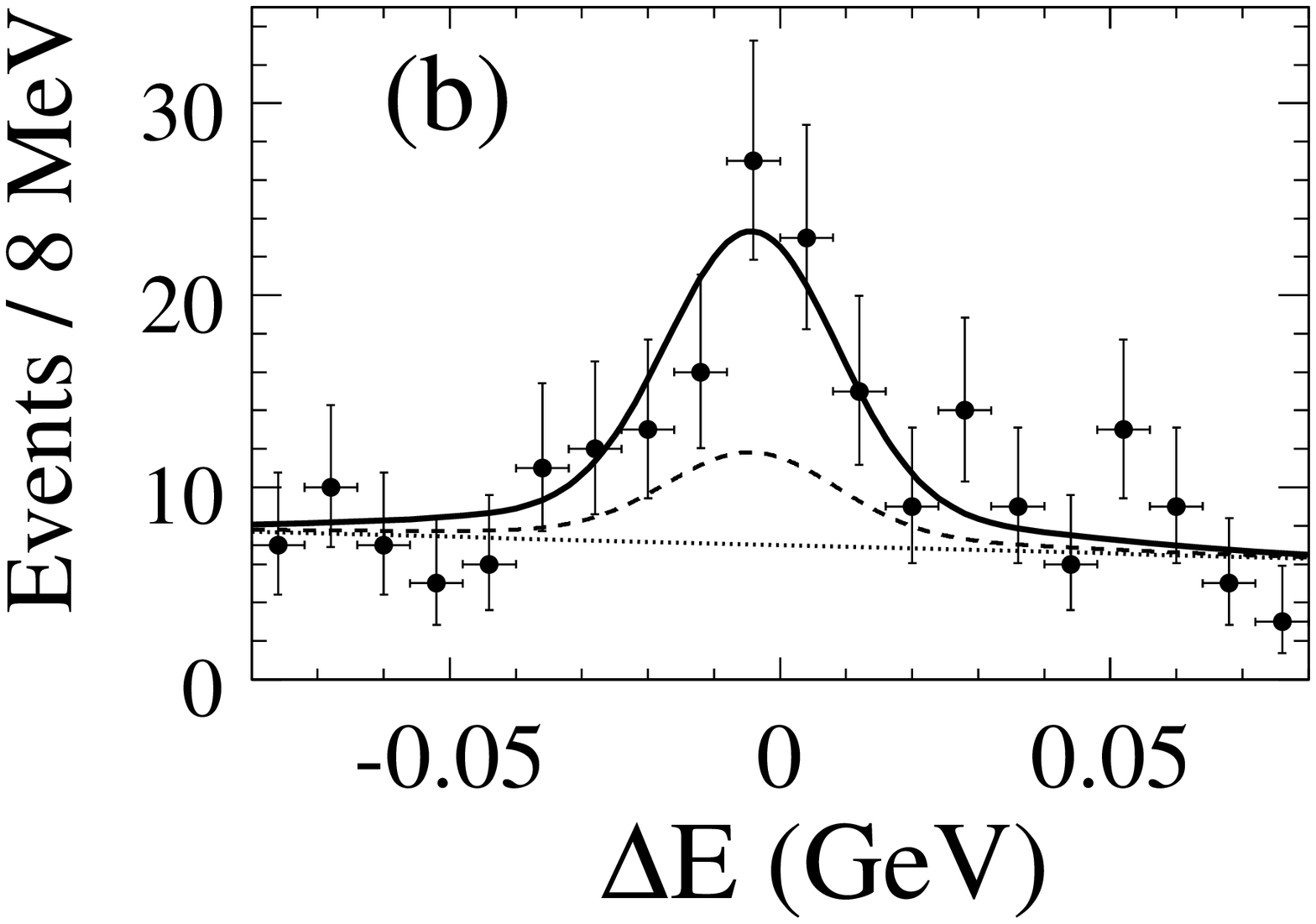}
\setlength{\epsfxsize}{0.5\linewidth}\leavevmode\epsfbox{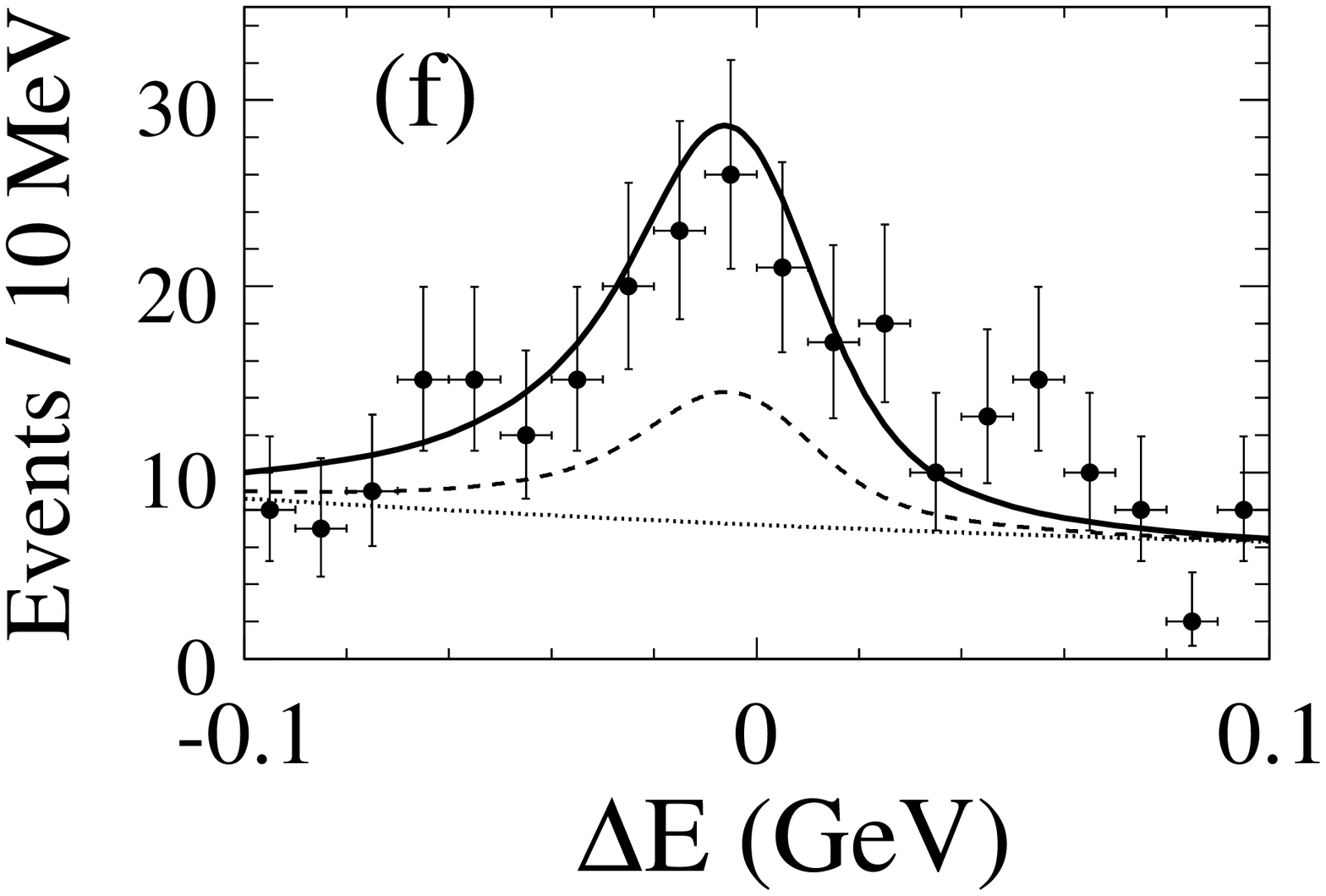}
}
\centerline{
\setlength{\epsfxsize}{0.5\linewidth}\leavevmode\epsfbox{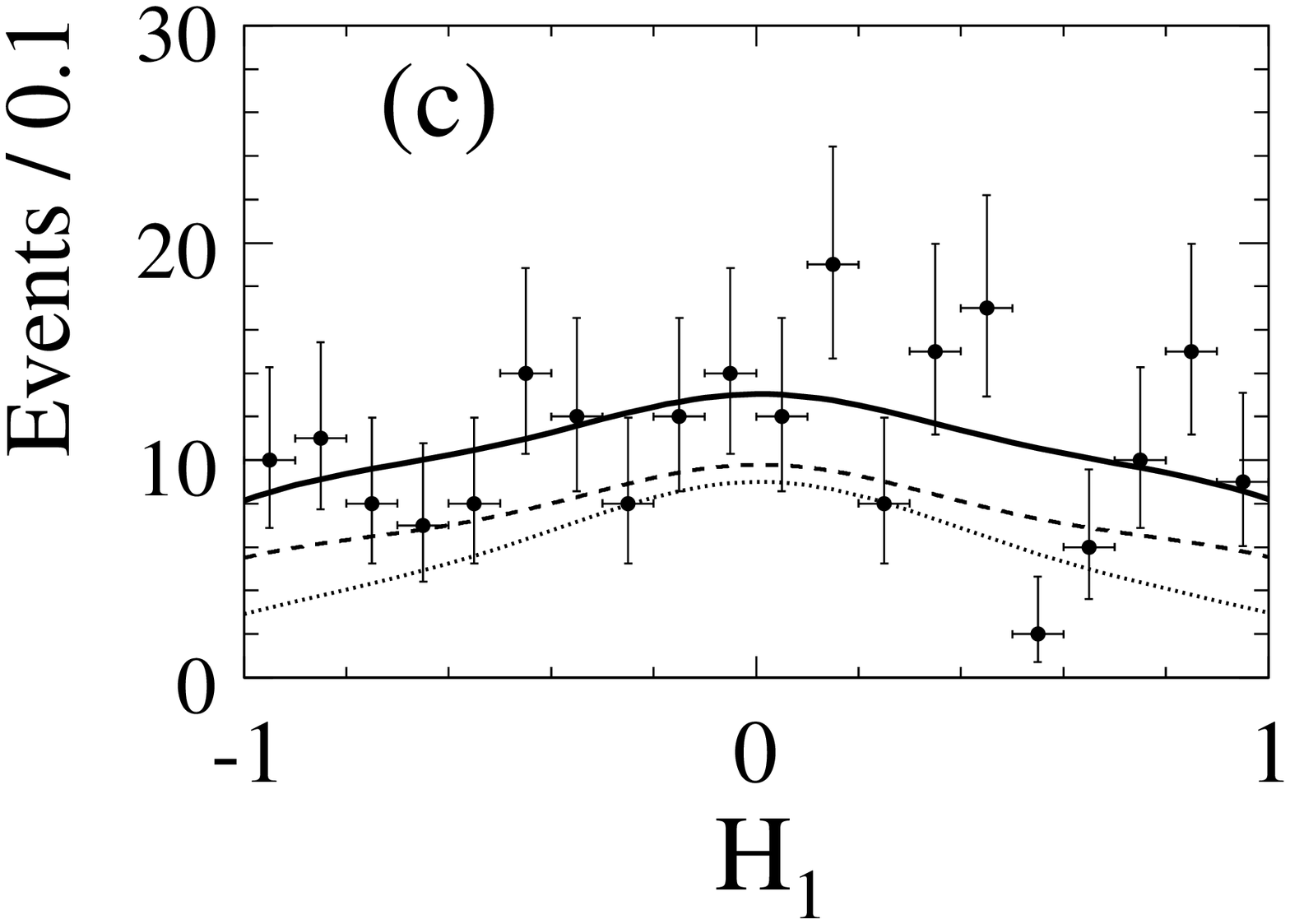}
\setlength{\epsfxsize}{0.5\linewidth}\leavevmode\epsfbox{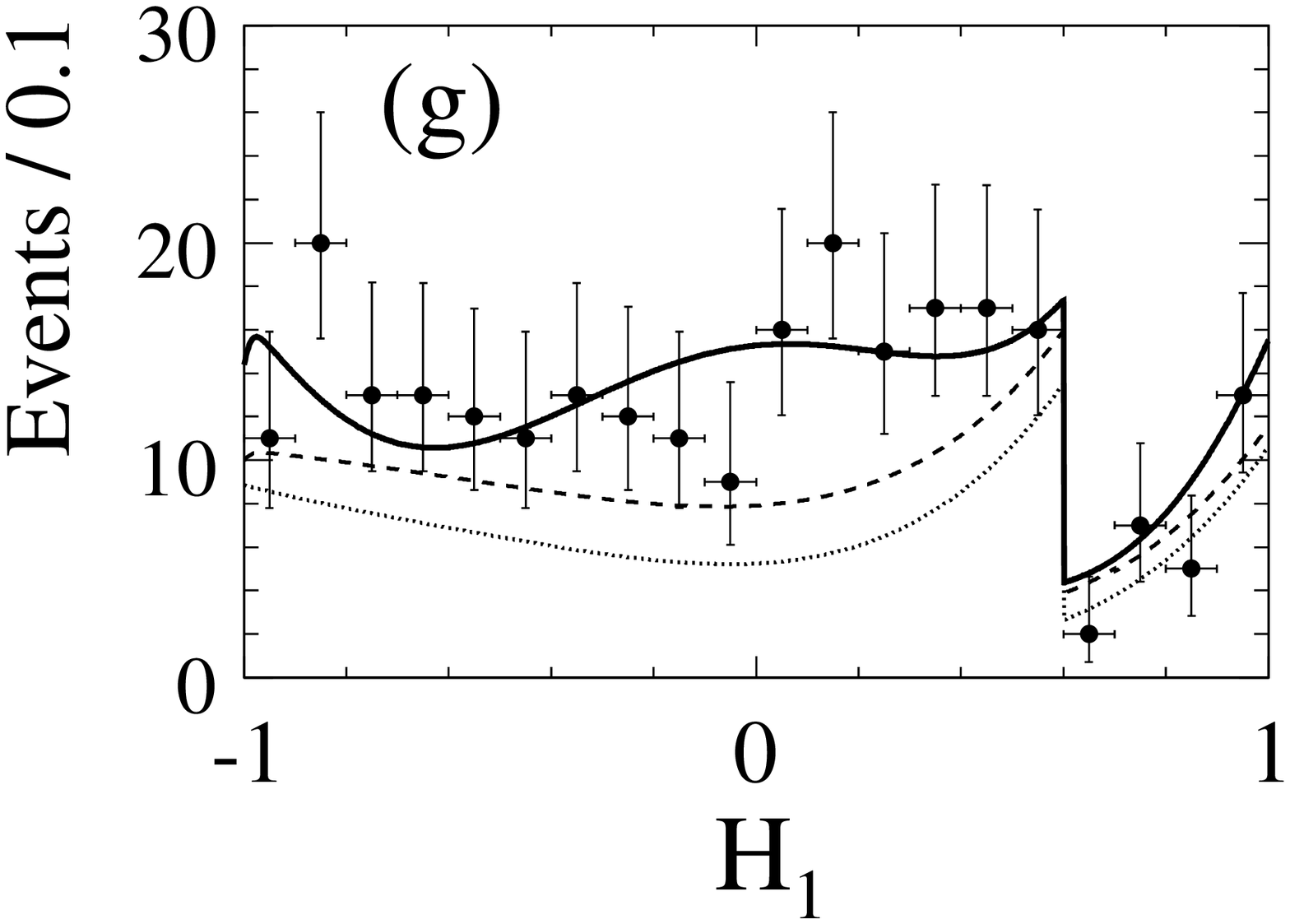}
}
\centerline{
\setlength{\epsfxsize}{0.5\linewidth}\leavevmode\epsfbox{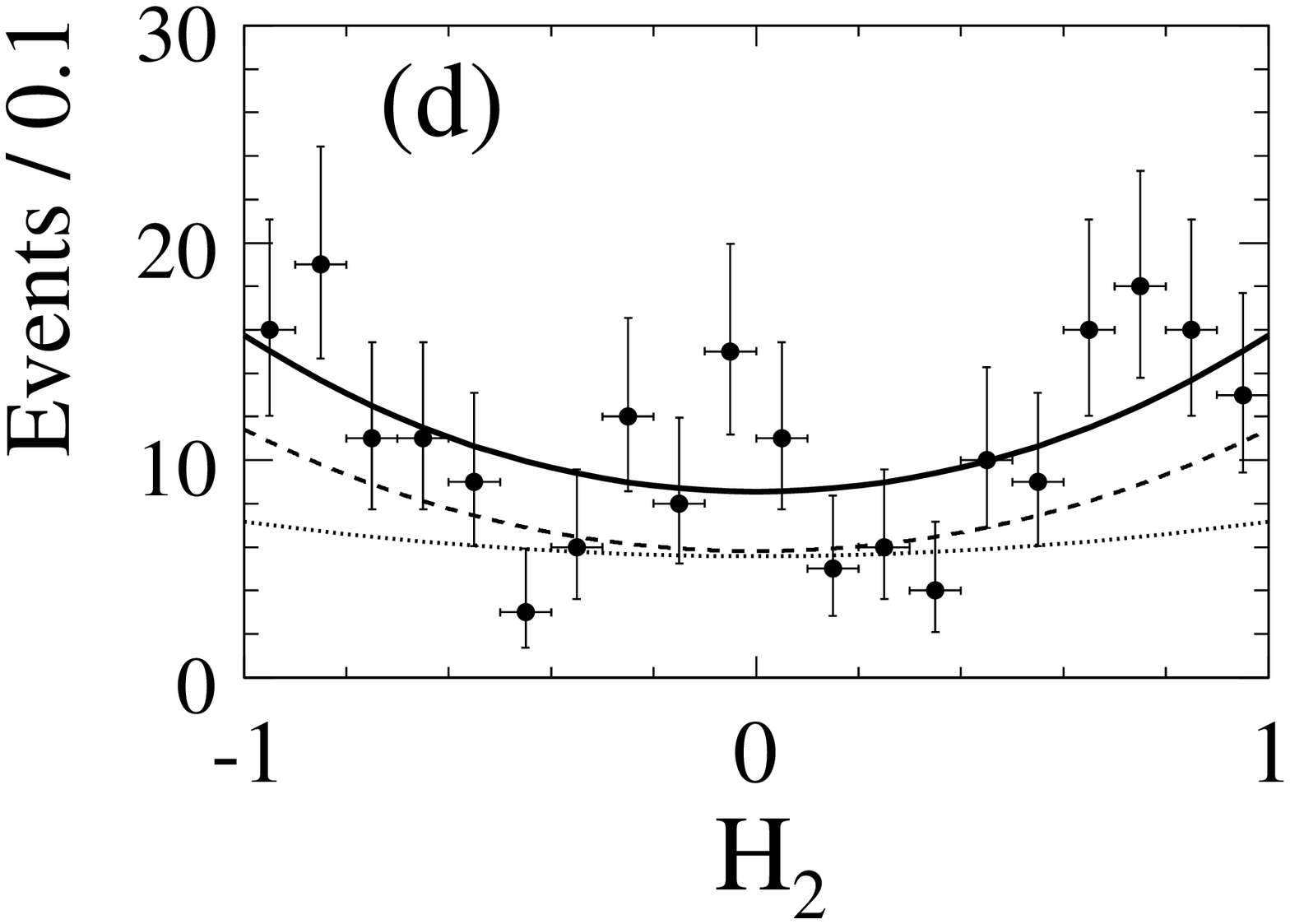}
\setlength{\epsfxsize}{0.5\linewidth}\leavevmode\epsfbox{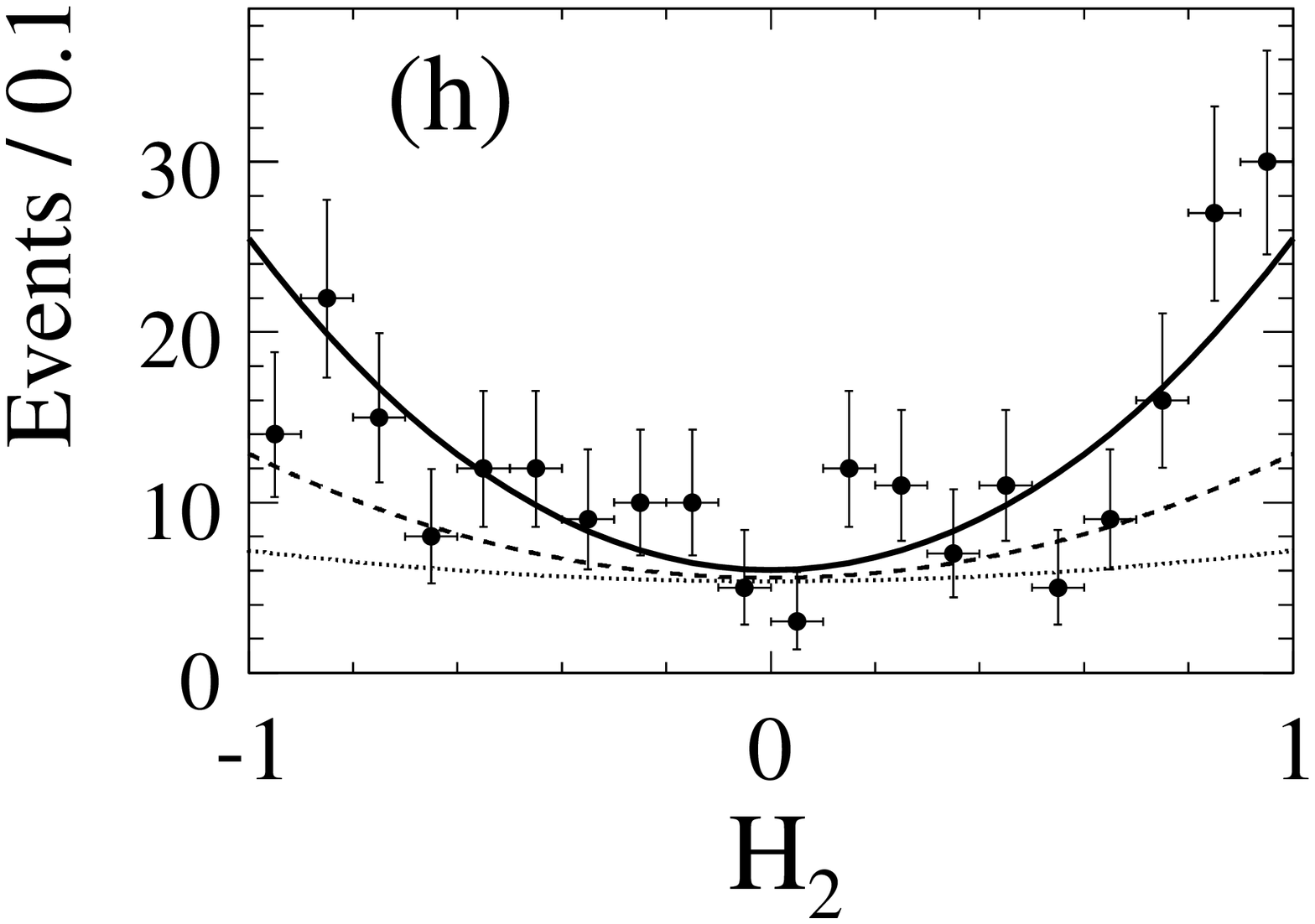}
}
\vspace{-0.3cm}
\caption{\label{fig:projection2} 
Left column: projections onto the variables  $m_{K\!\pi\!\pi}$ (a), $\Delta E$ (b), 
 ${\cal H}_1$ (c), and  ${\cal H}_2$ (d) for the signal $\varphi K_1(1270)^{\pm}$ candidate.
Right column: projections onto the variables  $m_{K\!\pi}$ (e), $\Delta E$ (f), 
 ${\cal H}_1$ (g), and  ${\cal H}_2$ (h) for the signal $\varphi K_2^{*}(1430)^{\pm}$ and
$\varphi (K\pi)_0^{*\pm}$ candidates combined.
The step in (g) is due to selection requirement ${\cal H}_1<0.6$ 
in the channel with $\pi^0$.
Data distributions are shown with a requirement on the signal-to-background
probability ratio calculated with the plotted variable excluded.
The solid (dotted) lines show the signal-plus-background
(combinatorial background) PDF projections, while the dashed lines 
show the full PDF projections excluding $\varphi K_1^{\pm}$ (left) or 
$\varphi K_2^*(1430)^{\pm}$ (right).
}
\end{figure}


We observe nonzero $B^\pm\to\varphi K_1(1270)^\pm$ 
and $B^\pm\to\varphi K^{*}_2(1430)^\pm$ 
yields with significance
(excluding systematic uncertainties in parentheses)
of 5.0(5.3)$\sigma$ and 5.5(6.0)$\sigma$, respectively.
The combined $\varphi K_1(1270)^\pm$ and $\varphi K_1(1400)^\pm$ 
significance is 5.7(6.4)$\sigma$.
The significance is defined as the square root of the change in 
$2\ln{\cal L}$ when the yield is constrained to zero in the 
likelihood ${\cal L}$.
We have tested this procedure with the generated MC samples
and account for a small observed deviation from the 
one-dimensional $\chi^2$ statistical treatment.

In Table~\ref{tab:results}, results of the fit are presented,
where the combined results are obtained from the simultaneous fit
to the three decay subchannels.
In the branching fraction calculations we assume
$K_2\to K_2^{*}(1430)\pi$ and 
${\cal B}(K^{*}(1410)\to K^{*}\pi)= 0.934\pm0.013$~\cite{bib:Amsler2008}.
The signal is illustrated in the projection plots in
Figs.~\ref{fig:projection1} and~\ref{fig:projection2},
where in the latter we enhance either the $\varphi K_1(1270)^\pm$
signal (left) or the $\varphi K^{*}_2(1430)^\pm$ signal (right).
The nonresonant $K^+K^-$ contribution under the
$\varphi$ is accounted for with the $B^0\to f_0 K_1$ category
and its yield  $7\pm 16$ is consistent with zero.
Similarly, the nonresonant category $\varphi K\pi\pi$ 
yield is $148\pm 54$ with statistical errors only.

\begingroup
\begin{table*}[t]
\caption
{\label{tab:results}
Results:
the reconstruction efficiency $\varepsilon_{\rm reco}$;
the total efficiency $\varepsilon$, 
including the daughter branching fractions~\cite{bib:Amsler2008}; 
the number of signal events $n_{\rm sig}$;  
significance ${\cal S}$; 
fraction of longitudinal polarization $f_L$;
the branching fraction ${\cal B}$;
and the flavor asymmetry ${\cal A}_{\CP}$.
The branching fraction ${\cal B}(B^\pm\to\varphi (K\pi)^{*\pm}_0$) refers to the coherent 
sum $|A_\text{res}+A_\text{non-res}|^2$
of resonant and nonresonant $J^P=0^+$ $K\pi$ components~\cite{Aston:1987ir} and 
is quoted for $m_{K\!\pi}<1.6$~GeV, while the ${\cal B}(B^\pm\to\varphi K_0^{*}(1430)^\pm)$ 
is derived from it by integrating separately the Breit-Wigner formula of the 
resonant $|A_\text{res}|^2$
$K\pi$ component~\cite{Aston:1987ir} without $m_{K\!\pi}$ restriction.
When several subchannels contribute, yield and efficiency are quoted 
for each subchannel. 
The 90\% confidence level upper limit on ${\cal B}$ is quoted
with the central values and errors in parentheses.
The insert shows two interference parameters $\delta_0$
and $\Delta\delta_0$ for  $\varphi K_2^{*}(1430)^\pm$ and $\varphi (K\pi)^{*\pm}_0$.
The $\varphi K_2(1770)^\pm$ yield 
is not considered in the nominal fit and the value indicated with ${\dagger}$
is obtained with these $\varphi K_2(1820)^\pm$ yield constrained to zero.
The systematic errors are quoted last.
}
\begin{center}
\begin{ruledtabular}
\setlength{\extrarowheight}{1.5pt}
\begin{tabular}{lccccccc}
\vspace{-3mm}&&&\\
Mode  
 & $\varepsilon_{\rm reco}$ (\%) & $\varepsilon$ (\%) & $n_{\rm sig}$ (events)
 & ${\cal S}$ ($\sigma$) & $f_L$ & ${\cal B}$  ($10^{-6}$) & ${\cal A}_{\CP}$ \cr
\vspace{-3mm} & & & & \\
\hline
\vspace{-3mm} & & & & \\
 $\varphi K_1(1270)^\pm$ 
 & $25.4\pm{1.4}$ & $4.07\pm 0.51$  & $116\pm 26~^{+15}_{-14}$
 & $5.0$ & $0.46^{+0.12}_{-0.13}~^{+0.06}_{-0.07}$ & $6.1\pm 1.6\pm 1.1$ & $+0.15\pm{0.19}\pm 0.05$  
\\ 
\vspace{-3mm} & & & &\\
 $\varphi K_1(1400)^\pm$ 
 & $24.6\pm{1.3}$ & $5.19\pm 0.44$  & $7\pm 39\pm 18$
 & $0.2$ &  & $<3.2$ ($0.3\pm{1.6}\pm 0.7$) &   
\\
\vspace{-3mm} & & & &\\
 $\varphi K_2^{*}(1430)^\pm$
 &   & $3.34\pm 0.14$  & $130\pm{27}\pm{14}$
  & $5.5$ & $0.80^{+0.09}_{-0.10}\pm0.03$ & $8.4\pm{1.8}\pm{1.0}$ & $-0.23\pm{0.19}\pm 0.06$  
\\
 ~~$\to K^0_S\pi^\pm$  & $11.9\pm{0.6}$ & $0.64\pm{0.04}$  & $27\pm{6}\pm{3}$  &  &  
\\
 ~~$\to K^\pm\pi^0$    & $12.2\pm{0.7}$ & $1.00\pm{0.06}$  & $39\pm{8}\pm{4}$  &  &  
\multicolumn{3}{c}{\framebox{$\delta_0=3.59\pm 0.19\pm 0.12$~~~~~$\Delta\delta_0=-0.05\pm 0.19\pm 0.06$}}
\\
 ~~$\to K^\pm\pi^+\pi^-$ & $24.7\pm{1.3}$ & $1.68\pm{0.12}$ & $64\pm{14}\pm{7}$  &  &  &   
\\
\vspace{-3mm} & & & & \\
 $\varphi (K\pi)^{*\pm}_0$
 & & $3.33\pm 0.13$  & $128\pm21 \pm 12$
  & 8.2 &  & $8.3\pm{1.4}\pm 0.8$ & $+0.04\pm{0.15}\pm 0.04$  
\\
 ~~$\to K^0_S\pi^\pm$  & $10.9\pm{0.6}$ & $1.24\pm{0.07}$  & $48\pm{8}\pm 4$  &  &  &   
\\
 ~~$\to K^\pm\pi^0$    & $12.8\pm{0.7}$ & $2.09\pm{0.12}$  & $80\pm{13}\pm 8$  &  &  &   
\\
\vspace{-3mm} & & & & \\
 $\varphi K_0^{*}(1430)^\pm$ 
 &  &  &  & 
 &  & $7.0\pm{1.3}\pm 0.9$ & 
\\
\vspace{-3mm} & & & &\\
 $\varphi K^*(1410)^\pm$ 
 & $28.0\pm{2.2}$ & $5.71\pm 0.44$  & $64\pm31{}^{+20}_{-31}$
 & $<2$ &  & $<4.3$ ($2.4\pm{1.2}~{}^{+0.8}_{-1.2}$) &   
\\
\vspace{-3mm} & & & &\\
 $\varphi K_2(1770)^\pm$ 
 & $20.8\pm{1.4}$ & $2.27\pm 0.16$  & $(90\pm 32~{}^{+39}_{-46})^{\dagger}$
 & $<2$ &  & $<15.0$  &   
\\
\vspace{-3mm} & & & &\\
 $\varphi K_2(1820)^\pm$ 
 & $21.6\pm{1.5}$ & $2.35\pm 0.18$  & $122\pm 40^{+26}_{-83}$
 & $<2$ &  & $<16.3$  &   
\\
\vspace{-3mm} & & & &\\
\end{tabular}
\end{ruledtabular}
\end{center}
\end{table*}
\endgroup


We vary those parameters in {\boldmath$\xi$} not used
to model combinatorial background within their
uncertainties and derive the associated systematic errors.
Interference between the $K_1(1270)^{\pm}$ and 
$K_1(1400)^{\pm}$ is one of the dominant systematic 
uncertainties on both yields and is modeled with simulated samples.
We take the flavor-dependent reconstruction efficiency into account in 
the study of asymmetries.
The biases from the finite resolution of the angle measurement,
the dilution due to the presence of false combinations, and
other imperfections in the signal PDF model are estimated
with MC simulation. Additional systematic uncertainty originates
from possible $B$ background, where we estimate from MC simulation
that only a few events can fall in the signal region.

The $\varphi K_2(1770)^\pm$ yield is not considered in the nominal fit 
due to large correlation with $\varphi K_2(1820)^\pm$.
But we substitute the $K_2(1820)$ resonance for the 
$K_2(1770)$ resonance, and find consistent results.
The difference is accounted as a systematics uncertainty, 
while the yield of decay $B^{\pm}\to\varphi K_2(1770)^\pm$ 
is used to obtain its branching fraction.
We quote only upper limits on the
two branching fractions as their correlation is not accounted for
in the central values. For the $\varphi K_2$ and $\varphi K^*(1410)$ decays, 
we vary the longitudinal polarization fraction between 
0.5 and 0.93, and constrain it to 0.8 in the nominal fit.
Polarization variations are included in the branching fraction calculations.
We vary the kinematic parameter describing 
$K_J^{\pm}\to K^{\pm}\pi^+\pi^-$ decay
($r_{02}$ in Ref.~\cite{vtpaper}) for various 
partial waves of the quasi-two-body $K_2$ decay 
channels and take the largest variations as the systematic uncertainties.
The systematic uncertainties in efficiencies are dominated
by those in particle identification, track finding,
and $K^0_S$ and $\pi^0$ selection.
Other systematic effects arise from event-selection criteria,
$\varphi$ and $K^{(*)}_J$ branching fractions, and number of $B$ mesons.


In summary, we have performed an amplitude analysis and searched
for $C\!P$-violation with the $B^\pm\to\varphi K_J^{(*)\pm}$ decays
which include significant $K_1(1270)$ and $K_2^{*}(1430)$ contributions.
Our results are summarized in Table~\ref{tab:results}.
The polarization measurement in the vector--tensor $B^\pm$ decay is
consistent with our earlier measurement in the
$B^0\to\varphi K_2^{*}(1430)^0$ decay~\cite{babar:vv} and with 
the naive expectation of the longitudinal polarization dominance. 
However, our first measurement of polarization in a vector--axial-vector 
$B$ meson decay indicates a large fraction of transverse amplitude, similar
to polarization observed in the vector--vector final state 
$B\to\varphi K^{*}(892)$~\cite{babar:vv,belle:phikst}.
Both measurements indicate substantial $A_{1+1}$ (or still possible $A_{1-1}$
for vector--axial-vector decay) amplitude from an uncertain source.
Among potential sources are penguin annihilation, electroweak penguin,
QCD rescattering, or physics beyond the standard model~\cite{smtheory}.


We are grateful for the excellent luminosity and machine conditions
provided by our \pep2\ colleagues,
and for the substantial dedicated effort from
the computing organizations that support \babar.
The collaborating institutions wish to thank
SLAC for its support and kind hospitality.
This work is supported by
DOE
and NSF (USA),
NSERC (Canada),
CEA and
CNRS-IN2P3
(France),
BMBF and DFG
(Germany),
INFN (Italy),
FOM (The Netherlands),
NFR (Norway),
MES (Russia),
MEC (Spain), and
STFC (United Kingdom).
Individuals have received support from the
Marie Curie EIF (European Union) and
the A.~P.~Sloan Foundation.


\end{document}